\documentclass[acmlarge,nonacm]{acmart}

\copyrightyear{2026}
\acmYear{2026}
\setcopyright{cc}
\setcctype{by-nc-nd}
\begin{document}

\title{Experiencing the More-than-Human Through Human Augmentation}

\author{Botao `Amber' Hu}
\orcid{0000-0002-4504-0941}
\affiliation{
  \institution{Reality Design Lab}
  \city{New York City}
  \country{USA}
}
\affiliation{
  \institution{University of Oxford}
  \city{Oxford}
  \country{UK}
}
\email{botao@reality.design}

\author{Danlin Huang}
\orcid{0009-0009-7954-0803}
\affiliation{%
    \institution{China Academy of Art}
    \city{Hangzhou}
    \country{China}
}
\email{danlinhuang0428@gmail.com}

\begin{abstract}
The recent more-than-human turn in design calls for attentiveness to nonhuman beings. Yet---as Thomas Nagel's famous ``What is it like to be a bat?'' thought experiment highlights---human experience is constrained by our own sensorium and an irreducible gap in phenomenal access to nonhuman \emph{Umwelten}. Grounded in eco-phenomenology and eco-somatics, this paper proposes \textbf{Experiencing the More-than-Human through Human Augmentation} (MtHtHA, or ``>HtH+''), a design approach that repurposes human augmentation technologies---typically aimed at enhancing human capabilities for human optimization---to create temporary, embodied, first-person experiences that modulate the human sensorium to approximate nonhuman sensory experiences, cultivating ecological awareness, empathy, and care across species boundaries. We articulate seven design principles, report five design cases---EchoVision (bat-like echolocation), FeltSight (star-nosed-mole tactile navigation), FungiSync (fungal network attunement), TentacUs (octopus-like distributed agency), and City of Sparkles (urban data from an AI's perspective)---and discuss implications for more-than-human aesthetics and design practice.
\end{abstract}


\begin{teaserfigure}
    \centering
    \includegraphics[width=1\linewidth]{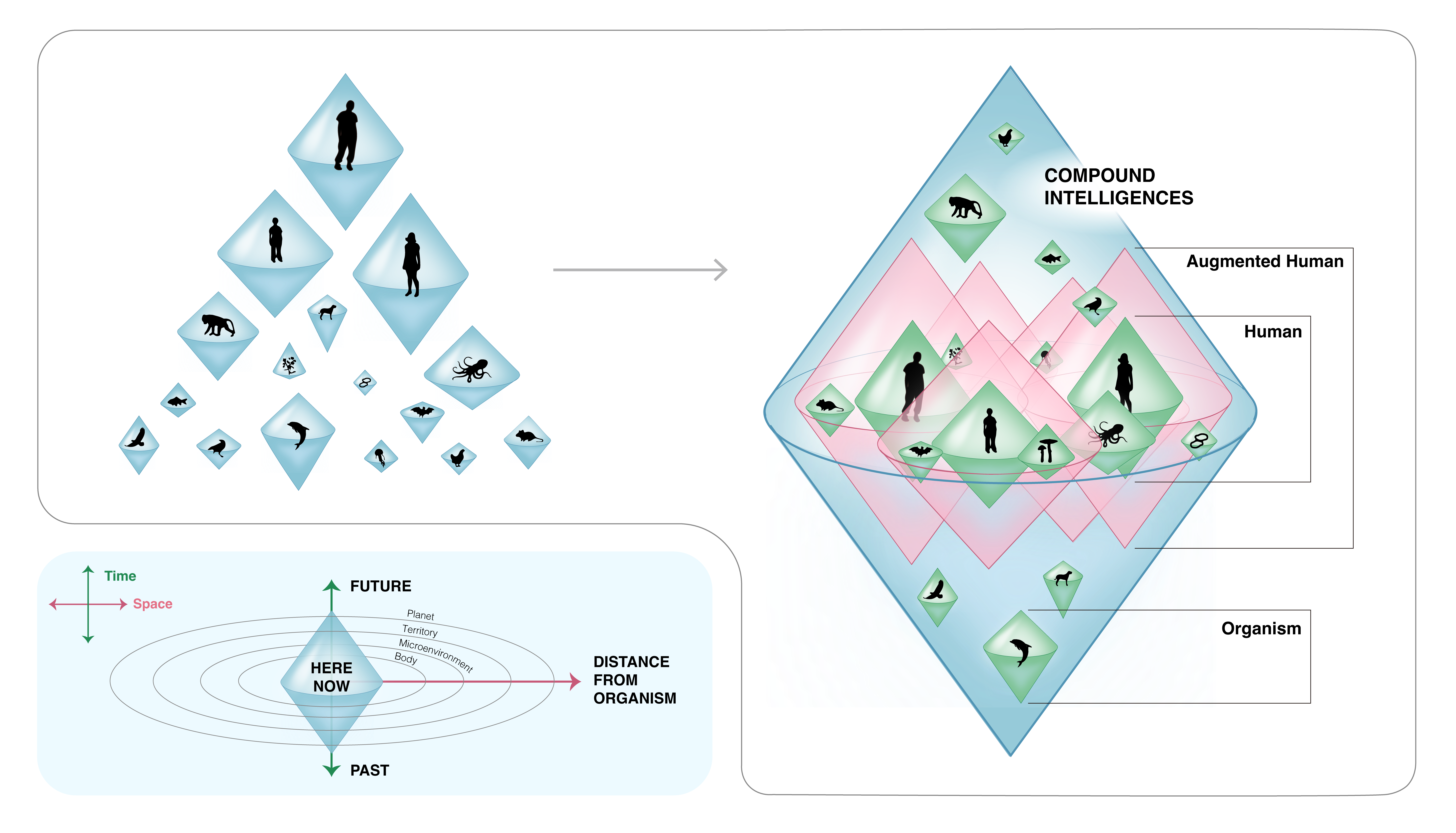}
    \caption{Experiencing the More-than-Human through Human Augmentation (MtHtHA) is a design approach that uses human augmentation technologies to create temporary, embodied, first-person experiences that approximate more-than-human Umwelten, cultivating ecological awareness, empathy, and care across species boundaries.}
    \label{fig:teaser}
\end{teaserfigure}

\maketitle

\section{Introduction}
The field of design has increasingly moved beyond human-centric optimization to embrace \emph{more-than-human} (MtH) perspectives, recognizing that humans exist in entangled ecologies, ``becoming-with'' other species, technologies, and environments \cite{Wakkary2021Thingsa,Coskun2022Morethanhuman,Giaccardi2020Technology,Abram1996spell,Escobar2018Designs}. This \emph{more-than-human turn} challenges designers to decenter from designing \emph{for} human use toward designing \emph{with} nonhuman actors---animals, plants, fungi, ecosystems, even microbes and AI---as partners or stakeholders \cite{Nicenboim2024MoreThanHuman,Nicenboim2025Decentering}.

However, a stubborn epistemic quandary complicates these efforts: an irreducible gap in phenomenal access to nonhuman \emph{Umwelten}---the self-worlds defined by each organism's unique sensory-motor apparatus \cite{Uexkull2010Foray}. As philosopher Thomas Nagel famously asked, \textit{``What is it like to be a bat?''} \cite{Nagel1974What}. Human experience is constrained by our own sensorium, and the subjective lifeworlds of nonhumans remain ultimately inaccessible to us. \emph{Eco-phenomenology} \cite{Brown2003Ecophenomenology} treats this gap not as a barrier but as a generative invitation to \emph{wonder} and \emph{empathize}: rather than replicating nonhuman experience (an impossibility), we can stage \textit{provisional, embodied approximations} that cultivate what \citet{Haraway2016Staying} calls ``response-ability''---heightened ethical responsiveness and attentiveness toward other beings \cite{PuigdelaBellacasa2017Matters}. By \emph{empathy}, we mean not cognitive role-taking but \emph{embodied, somatic feeling-toward}: first-person bodily resonance registered in sensation and proprioception rather than propositions.

More-than-human aesthetics \cite{Wilkie2025aesthetics,Sehgal2024MoreThanHuman} has been explored through design artifacts. As \citet{Ikeya2025Aesthetics} observe, \emph{phenomenological} inquiry is a key critical orientation in more-than-human aesthetics, particularly in how artifacts mediate nonhuman sensory experience. These approaches detect and translate nonhuman responses through visual, auditory, and functional representations, but remain largely \emph{representational}---mediated through screens, speakers, or conceptual tools---rather than \emph{somatic}. We aim to address this gap by employing embodied augmentation to restructure the human sensorium.

Meanwhile, \emph{human augmentation} (HA) encompasses wearables, AI, extended reality (XR), brain-computer interfaces, and biotech that extend human capabilities beyond their normal limits \cite{Raisamo2019Human}. Yet the dominant narrative of HA has pursued human enhancement with a transhumanist bent---envisioning humans as infinitely improvable \cite{More2013transhumanist}. Rather than pursuing superhuman optimization, we introduce \textbf{Experiencing the More-than-Human through Human Augmentation} (MtHtHA, or ``>HtH+''), a design approach that repurposes HA technologies toward eco-phenomenological goals. MtHtHA stages temporary, first-person bodily encounters through somatic resensitization, sensory reconfiguration, and ritualized interactions that make interspecies relations tangible without claiming equivalence, drawing on speculative design \cite{dunne2024speculative}, somaesthetic design \cite{Hook2018Designingd,Shusterman2007Body}, and eco-somatics \cite{Kuppers2022Eco,Kampe2021Embodying}.

This paper contributes three things. First, we articulate the MtHtHA approach and seven design principles for crafting augmentations that foster felt awareness, eco-somatic empathy, and care beyond the human. Second, we present five design cases: \emph{EchoVision} (bat-like echolocation), \emph{FeltSight} (star-nosed-mole tactile navigation), \emph{FungiSync} (mycelial perspective-sharing), \emph{TentacUs} (octopus-like distributed agency), and \emph{City of Sparkles} (urban data from an AI's perspective). Third, we discuss implications for expanding cognitive light cones \cite{Levin2019Computationalb} for understanding more-than-human worlds, the role of somatic defamiliarization \cite{Benford2012Uncomfortableb}, and opportunities for cultivating eco-soma literacy in design practice \cite{PoikolainenRosen2025Morethanhumana} and education \cite{Nilsson2025Navigating}.

\section{Background}

We ground MtHtHA in three intersecting lineages---more-than-human design, eco-phenomenology (and eco-somatics), and human augmentation research---then situate our approach within the emerging landscape of more-than-human aesthetics. Rather than surveying the full breadth of MtH scholarship \cite{Coskun2022Morethanhuman,PoikolainenRosen2025Morethanhumana,Nicenboim2024MoreThanHuman}, we focus on the strands most relevant to somatic augmentation and felt bodily encounter.

\subsection{More-than-Human Design and Eco-Phenomenology}

The term more-than-human, popularized by \citet{Abram1996spell}, rejects human exceptionalism and foregrounds co-dependence with animals, plants, materials, and technologies. In HCI and design, this has catalyzed a ``more-than-human turn'' \cite{Escobar2018Designs, Giaccardi2020Technology, Coskun2022Morethanhuman, Wakkary2021Thingsa, PoikolainenRosen2025Morethanhumana} drawing on STS \cite{Latour1993We,Bennett2010Vibrant,Barad2006Meeting}, multispecies anthropology \cite{Kohn2013How,Tsing2015Mushroom}, and environmental philosophy \cite{Hoffman2005Gettinga}. \citet{Haraway2016Staying}'s call to ``stay with the trouble'' and cultivate response-ability has inspired approaches treating nonhumans as co-designers or co-inhabitants \cite{Giaccardi2020Technology,Loh2024morethanhuman}. Recent work on more-than-human design practice and teaching more-than-human perspectives in HCI and design education highlights both opportunities and challenges of translating these theories into concrete design methods \cite{PoikolainenRosen2025Morethanhumana, Nilsson2024Teaching}.

Eco-phenomenology extends phenomenology to ecological relations by emphasizing that perception is always embodied, situated, and co-constituted with the environment \cite{Brown2003Ecophenomenology}. \citet{Merleau-Ponty2013Phenomenologyb} framed perception as a dialogue between body and world; \citet{Uexkull2010Foray}'s Umwelt concept highlights that each species inhabits its own sensory world defined by its perceptual and motor capacities. This has profound implications for design: humans can never fully access another species' lived experience, yet we can still seek partial, respectful attunements. Postphenomenology provides theoretical grounding for MtHtHA: \citet{Ihde1990Technology} showed how technologies restructure perception, while \citet{Verbeek2005WhatThings}'s mediation theory tracks how artifacts shape what we perceive and feel. Related work on interspecies play \cite{Westerlaken2016Playful} and virtual environments for experiencing nonhuman modes of being \cite{Gualeni2014Augmented,GualeniWesterlaken2013Zoomorphism} further demonstrates that designed encounters can foster perceptual shifts toward more-than-human others. Design scholarship has also translated these ideas into practices of decentering the human and designing for pluriversal worlds that acknowledge multiple ontologies and agencies \cite{Wright2014Becomingwith,Nicenboim2025Decentering,Nicenboim2024MoreThanHuman}.

\subsection{Human Augmentation versus Transhumanism}

Human augmentation (HA) encompasses technologies such as wearables, XR, haptics, brain-computer interfaces, and AI companions that extend human capabilities \cite{Raisamo2019Human}---from \citet{engelbart2023augmenting}'s vision of computer systems as tools to ``augment human intellect'', to \citet{Weiser1993Hot}'s ubiquitous computing and contemporary work on human-centric intelligence augmentation, to recent cyborg psychology \cite{pataranutaporn2024cyborg}. Such augmentation often aims to improve productivity, safety, or accessibility, remaining relatively aligned with normative human-centered goals \cite{Guerrero2022Augmented}. Transhumanism presents a more radical project: leveraging biotechnology and AI to fundamentally transcend biological limitations and ultimately re-engineer the human species \cite{More2013transhumanist,Hansell2011transhumanism}. While HA and transhumanism share a focus on enhancement, they diverge in ethos. As \citet{Rakkolainen2026Augmented} notes, HA usually seeks to empower people as they are, whereas transhumanists push to redesign humans entirely. Critics observe that enhancement discourses frequently reproduce anthropocentric, individualistic assumptions---privileging optimization over care, control over reciprocity \cite{Haraway2016Staying,Barad2006Meeting}. More-than-human design scholarship argues that technologies should support living well together with other species rather than solely maximizing human capacities \cite{Escobar2018Designs,Giaccardi2020Technology}.

MtHtHA aligns with this critical reorientation: instead of pursuing superhuman abilities, we reshape the human sensorium to attune to other beings. Rather than treating bodily limits as defects to overcome, we treat them as design materials that can be reconfigured to foreground entanglements with more-than-human others---reframing HA from a project of human supremacy to one of multispecies sensitivity \cite{PuigdelaBellacasa2017Matters}.

\subsection{Embodied Interaction, Somaesthetics, and Eco-somatics}

Embodied interaction emphasizes that meaning arises through bodily engagement with technology and environment rather than symbolic information processing \cite{Dourish2004Where}. \citet{Shusterman2007Body}'s somaesthetics proposes both a philosophical account and a practical program for improving bodily awareness, which \citet{Hook2018Designingd} translated into soma design methods---guided bodily explorations, movement exercises, and felt-based prototyping---positioning the designer and user as sensing, responsive soma whose subtle feelings become central design materials \cite{Hook2016Somaesthetica}.

Eco-somatics extends somaesthetic concerns to explicitly ecological relationships. \citet{Kuppers2022Eco}'s notion of eco-soma encourages practitioners to feel their bodies as porous and entangled with land, weather, plants, and animals \cite{Kampe2021Embodying}. Eco-somatic practices often draw on feminist and Indigenous epistemologies that reject dualisms between mind and body or human and nature \cite{Haraway2016Staying,Hook2021Unpackinga}. MtHtHA builds directly on these traditions: our principles prioritize first-person felt experience, ritualized bodily engagement, and somatic resensitization as means to encounter more-than-human worlds.

\subsection{More-than-Human Aesthetics and Design}

More-than-human aesthetics \cite{Wilkie2025aesthetics,Sehgal2024MoreThanHuman} has been discussed and practiced for years \cite{PoikolainenRosen2025Morethanhumana,Giaccardi2025makings}. \citet{Ikeya2025Aesthetics} provide a comprehensive taxonomy of 83 artifacts of more-than-human design and identify three critical orientations: (1) \emph{phenomenological}---mediating nonhuman sensory experience; (2) \emph{ontological}---embracing relational intra-actions that refuse the subject--object dichotomy; and (3) \emph{conceptual}---provoking posthuman ethico-political reflection. Their analysis shows that the phenomenological orientation, which mediates nonhuman sensory experience through visual, auditory, and functional representations, is the most populated dimension. While these artifacts translate nonhuman responses into human-perceivable forms---mediated through screens, speakers, or conceptual tools---few are embodied. MtHtHA extends beyond translation to somatic restructuring, reconfiguring the human body itself as the site of nonhuman perceptual encounter.

MtHtHA sits within a lineage of precedent design practices and artworks with similar aesthetics. In XR art, \emph{In the Eyes of the Animal} \cite{mlf2015iteota} uses volumetric capture and immersive sound to place participants in forest animals' perceptual worlds; GoatMan \cite{Thwaites2016GoatMan} involves a designer living among goats using custom prosthetics to approximate a goat's Umwelt. The VR piece \emph{Tree} \cite{newrealitycotree} embodies participants as a rainforest tree---from seedling to canopy---evoking grief during deforestation. \emph{Eye of Flora} \cite{Hu2024Eye} uses AR to reveal pollinator-visible ultraviolet patterns, echoing an ethos of plant-centered design \cite{Loh2024morethanhuman}. Cyborg performances like \citet{Ribas2020Waiting}'s seismic-sense implant extend perception into geophysical events, illustrating how augmentation can reorient the body toward Earth systems.

These practices suggest that defamiliarizing embodied perception can invite reflection on nonhuman lifeworlds, but many remain one-off art pieces without articulated frameworks to guide ethical, ecological, or somatic considerations. MtHtHA aims to distill these practices into a design approach that deepens the phenomenological orientation of more-than-human aesthetics.

\section{Experiencing the More-than-Human Through Human Augmentation: Principles and Approach}

Our MtHtHA design approach is guided by seven principles distilled from prior design work and our own iterative practice. These principles shaped each case study and can inform future designs that use human augmentation to foster more-than-human awareness. For clarity, we use \emph{EchoVision} \cite{Hu2024Becoming} as an example to illustrate these principles.

\paragraph{(P1) Embodied Umwelt Approximation}
Temporarily place the user in a reimagined sensory world of a nonhuman other. Use HA technology to simulate key aspects of a nonhuman being's Umwelt (its perceptual world) in an \emph{embodied, first-person} manner, optionally incorporating ritualized interactions. The goal is an eco-phenomenological attunement to more-than-human worlds---what \citet{Ikeya2025Aesthetics} term the ``phenomenological orientation'' of more-than-human aesthetics---guiding users toward an approximation sufficient to spark imagination and empathy, rather than an exact replica. For example, \emph{EchoVision} gives humans the experience of ``seeing through sound'' like a bat. Rather than using real ultrasound bouncing off surfaces, it translates human-audible vocal input into visual effects on terrain, allowing participants to experience the bat's Umwelt through the user's active vocalization. This principle allows users to experience another being's perceptual world through their own bodies, cultivating experiential feeling rather than mere intellectual knowing.

\paragraph{(P2) Invite Accessible Participation}
Experiences should be inviting, accessible, intriguing, engaging, and understandable to laypeople---not only experts. Lowering barriers to entry (both technical and comfort-related) while making the experience \textit{attractive, intuitive, provocative, and playful} encourages the general public to inhabit more-than-human perceptions and maximizes public awareness. In \textit{EchoVision}, visual hooks (the bat-shaped mask) attract bystanders; low-friction entry (a handlebar allows holding the mask instead of head-mounting); intuitive interactions (speaking to trigger echoes); and site-specific exhibition (at the largest bat colony in the US, Austin's Congress Avenue Bridge) ensure anyone can participate. This principle supports cultivating widespread ecological awareness through broad public engagement.

\paragraph{(P3) From Awareness to Empathy to Care}
The ultimate aim is not novelty but \emph{empathic concern}. Design to evoke an emotional trajectory: awareness → empathy → care. First, capture attention and create \emph{awareness} of the more-than-human, opening space for \emph{empathy}---affective perspective-taking and emotional connection---which in turn can motivate \emph{care}---an ethical stance and willingness to act responsibly. In EchoVision, public participants actively passed the mask from one person to the next at the bat-colony site, teaching each other how to use it and explaining echolocation after experiencing it themselves. This principle aims to transform a momentary experience into lasting attitude change, drawing on \citet{Tronto1993Moral}'s insight that caring \emph{for} something begins with caring \emph{about} it \cite{PuigdelaBellacasa2017Matters}.

\paragraph{(P4) Ground in Somaesthetic Experience} Prioritize felt first-person bodily experience over intellectual understanding. Participants should learn through feeling, not explanation, making sense through their senses. In \textit{EchoVision}, users are not told upfront that ``this simulates bat echolocation.'' One can intellectually know that bats use echolocation, but feeling one's own voice mapped to vision provides a tacit understanding that no verbal description can convey. Factual knowledge---such as knowing that bats echolocate---becomes more meaningful after experiencing echo-based navigation firsthand. This principle aligns with soma design and eco-soma's emphasis on embodied knowledge. Such ``knowledge-by-doing'' fosters deeper connection and a sense of wonder.

\paragraph{(P5) Effective Defamiliarization} Embrace a degree of defamiliarization, discomfort, or uncanniness as a design feature, within safe bounds. Research on uncomfortable interactions \cite{Benford2012Uncomfortableb} suggests that carefully challenging a user's comfort can yield reflection and insight. We design experiences that intentionally push users slightly beyond their habitual comfort zones to disrupt routine perception. However, we ensure psychological and physical safety: these discomforts are brief, explainable, and under the user's control (they can always opt out). In \textit{EchoVision}, standing under a public bridge while wearing a bat mask and vocalizing odd sounds is strange yet enchanting---sparking conversations among participants about bats and senses that would not occur in mundane settings. Drawing from performance art and critical design, this principle uses discomfort to prompt critical awareness, requiring a balance: the experience should be challenging enough to be transformative, but not so overwhelming that it deters engagement.

\paragraph{(P6) Sensory Reprioritization and Constraint}
Deliberately reprioritize human senses---often by diminishing a dominant sense---to open space for others. Human perception is famously vision-centric (ocularcentric), which can narrow our awareness. By limiting or altering vision, we encourage users to rely on touch, sound, or other senses, more closely resembling how many animals perceive. This sensory constraint slows users down and encourages mindfulness. By \emph{removing} our dominant sense, we allow other sensory pathways to emerge, akin to how sensory deprivation can heighten attention to remaining modalities. For example, participants in \textit{EchoVision} exploring a dark cave reported that after initial disorientation, they grew accustomed to actively vocalizing to ``see'' their surroundings. This principle reframes augmentation to mean \emph{subtraction and reordering}---a counterintuitive strategy that can reveal aspects of the environment normally overshadowed by vision or other dominant senses.

\paragraph{(P7) Cross-Modal Mapping} Translate nonhuman sensory inputs into forms accessible to humans. Many more-than-human signals---ultrasonic sound, electromagnetic fields, pheromones---are imperceptible to unaided humans. We use sensors and code as translation layers. For example, \emph{EchoVision} maps bouncing sound to visual effects over LiDAR-scanned terrain, effectively giving people an augmented sense they lack. This principle builds on sensory substitution research showing that the brain can learn to interpret novel sensory mappings \cite{Bach-Y-Rita1969Vision,Striem-Amit2012Reading,Amedi2007Shape}. By designing reasonable mappings, we let users learn a new sense on the fly. Even if the learning is rudimentary in a short session, it reveals our latent capacity to extend perception---a humbling experience that blurs ``natural'' limits.

These principles serve as intersecting design heuristics. Together, they form a toolkit for fostering what we call \textit{eco-somatic empathy}: a felt, bodily resonance that arises when the body is reconfigured to approximate another's perceptual situation \cite{Hook2018Designingd,Kuppers2022Eco}---registered in proprioception rather than propositions.

\section{Case Studies}

The following five case studies exemplify the MtHtHA approach and demonstrate these design principles in practice, each targeting a different nonhuman Umwelt.

\subsection{Becoming a Bat with ``EchoVision'': Seeing Through Sound}
\label{sub:echovision}

\begin{figure}[ht]
    \centering
    \includegraphics[width=1\linewidth]{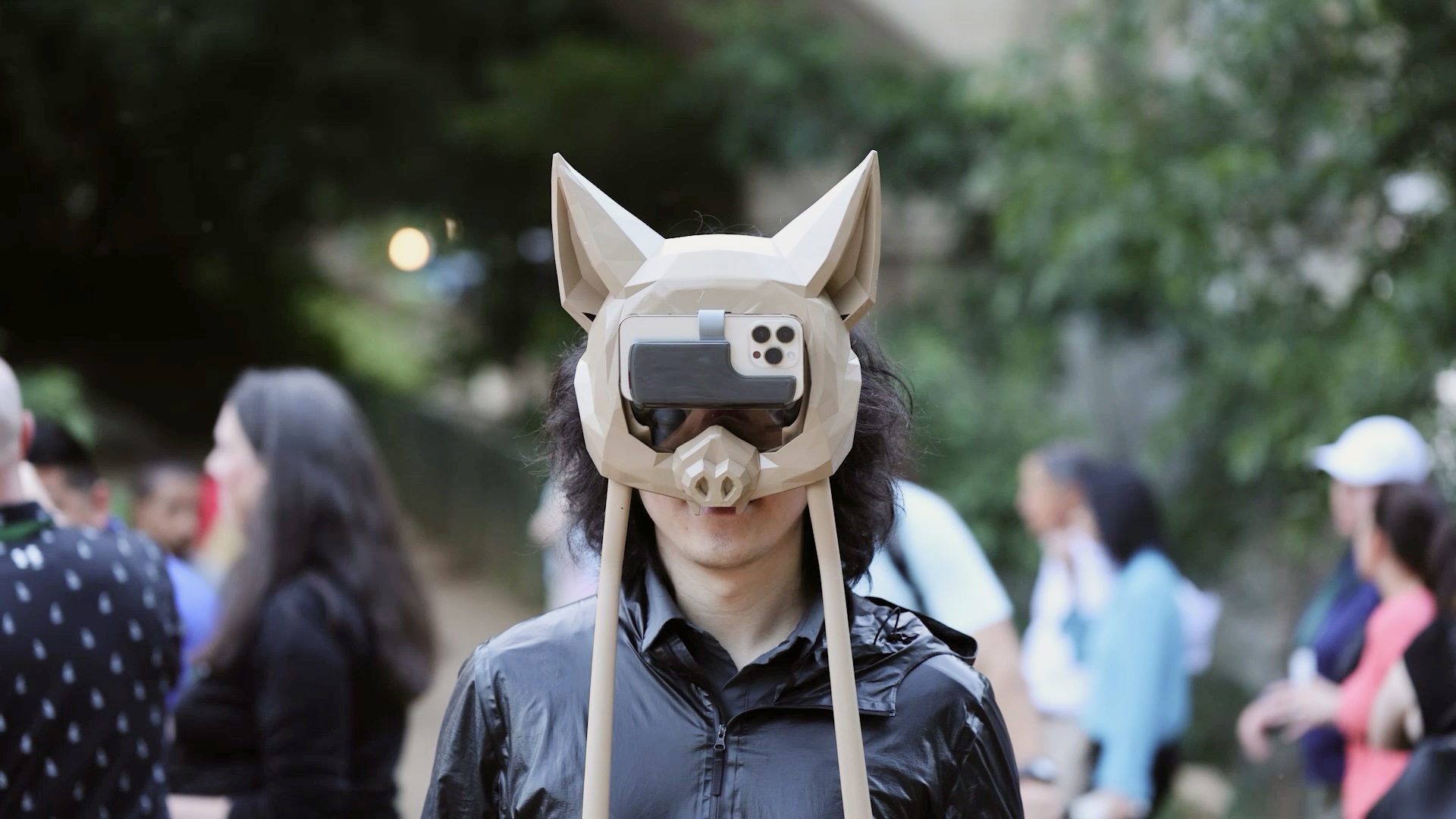}
    \caption{EchoVision is an immersive art experience using a mixed reality handheld mask to simulate bat echolocation.}
    \label{fig:echovision}
\end{figure}

\textit{EchoVision} \cite{Hu2024Becoming} directly responds to Nagel's question ``what is it like to be a bat?'' by offering participants a mixed-reality (MR) experience of bat echolocation. The project uses a custom bat-shaped mask that houses the open-source HoloKit MR headset \cite{Hu2024HoloKit}. When participants wear the mask and vocalize, the system maps sound waves to visual echoes in the MR display, allowing them to ``see'' the surrounding environment through reflected sound. The mask itself serves as a curiosity catalyst, inviting bystanders to imagine the unseen experience, thus fulfilling P2. In a site-specific pop-up under Austin's Congress Avenue Bridge---the habitat of millions of free-tailed bats---over two hundred participants experienced the installation in collaboration with bat conservationists.

\textit{EchoVision} intentionally estranges human perception (P5) by converting auditory information into visual cues, encouraging participants to explore without relying on sight. Participants described moving more slowly, listening intently, and appreciating the presence of bats above them. By fostering an embodied awareness of echolocation, the piece transforms abstract knowledge about bats into a felt experience, aligning with P4. The shared, public nature of the installation also created interspecies dialogue; participants vocalized to detect each other, while bystanders observed and photographed the interactions, creating a collective ecological awareness.

\subsection{Becoming a Star-Nosed Mole with ``FeltSight'': Touching Beyond Reach}

\begin{figure}[ht]
    \centering
    \includegraphics[width=1\linewidth]{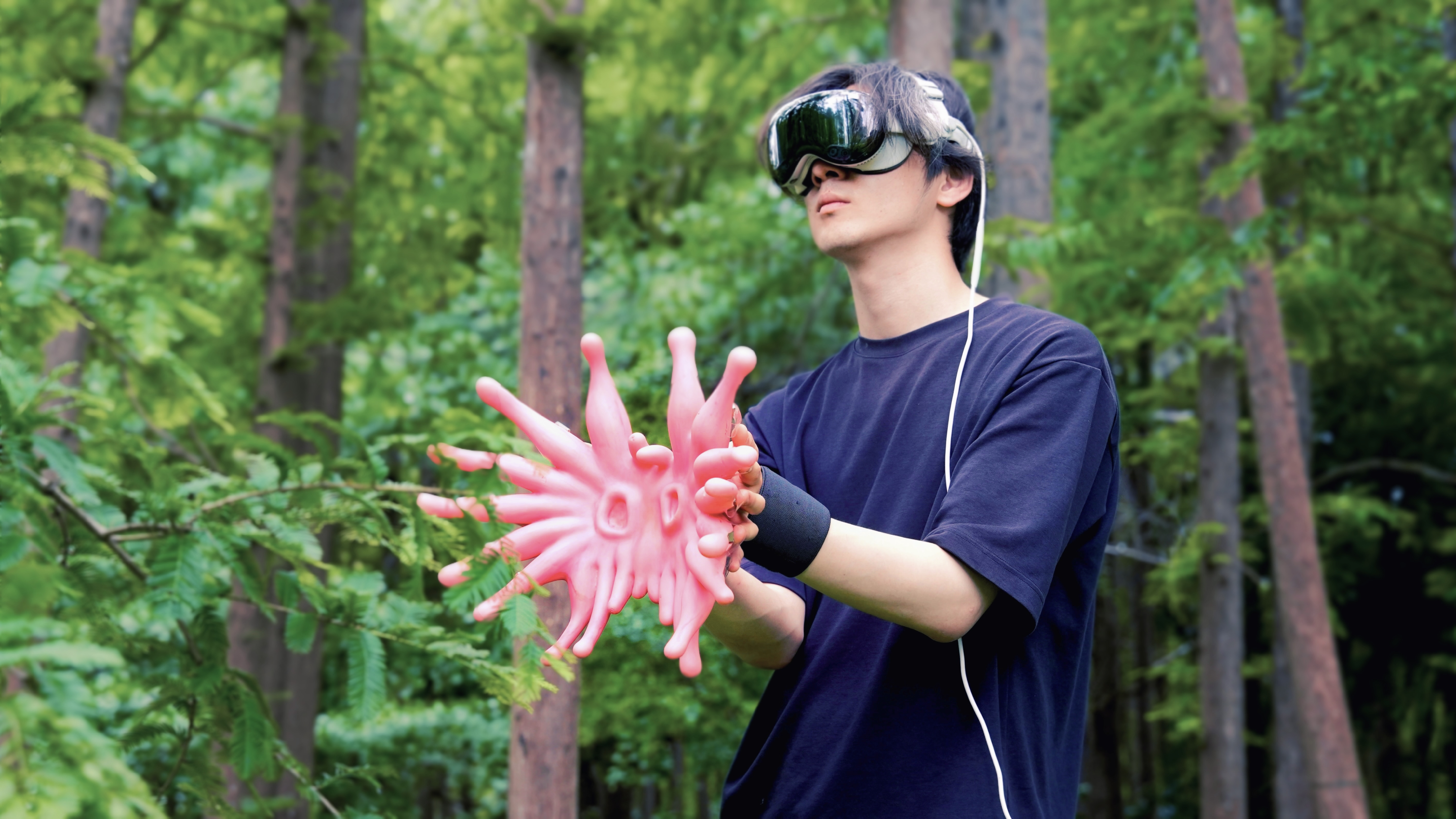}
    \caption{FeltSight is a star-nosed-mole-inspired system that combines custom haptic gloves with a mixed reality headset to extend tactile perception beyond the skin, allowing users to ``feel'' material textures of surrounding objects without physical contact.}
    \label{fig:feltsight}
\end{figure}

Humans are profoundly ocularcentric---a cultural and physiological prioritization of vision over other senses. \emph{FeltSight} \cite{Huang2025Felt} challenges this bias by diminishing visual input and elevating touch. Inspired by the star-nosed mole, an animal that is functionally blind yet perceives through a highly sensitive tactile organ containing over 100,000 sensory receptors, FeltSight comprises a haptic glove with vibrotactile actuators paired with an XR headset. When users point toward nearby objects, LiDAR sensing and haptic feedback allow them to feel textures without physical contact. The visual display shows only a minimal point cloud representing ``felt memory,'' compelling participants to navigate through tactile exploration.

Participants reported an initial sensory rupture: the darkness of the headset and absence of visual landmarks induced uncertainty and loss of control, aligning with P5's notion of productive discomfort. Gradually, they adopted slow, exploratory movements reminiscent of burrowing animals and described an extended body schema, feeling as if their fingers had elongated beyond their physical reach. This resonates with Merleau-Ponty's view of perception as an active interplay between body and environment and illustrates how re-prioritizing senses (P6) can open new modes of attention. A meditation practitioner noted that the experience offered a novel understanding of walking meditation, suggesting that our vision-dominated cognition can impede flow states. FeltSight thus encourages participants to attune to subtle textures, promoting mindful engagement and empathy for creatures that navigate through touch.

\subsection{Entangling Like Mycorrhizae with ``FungiSync'': Mixing Realities Through Touch}

\begin{figure}[ht]
    \centering
    \includegraphics[width=1\linewidth]{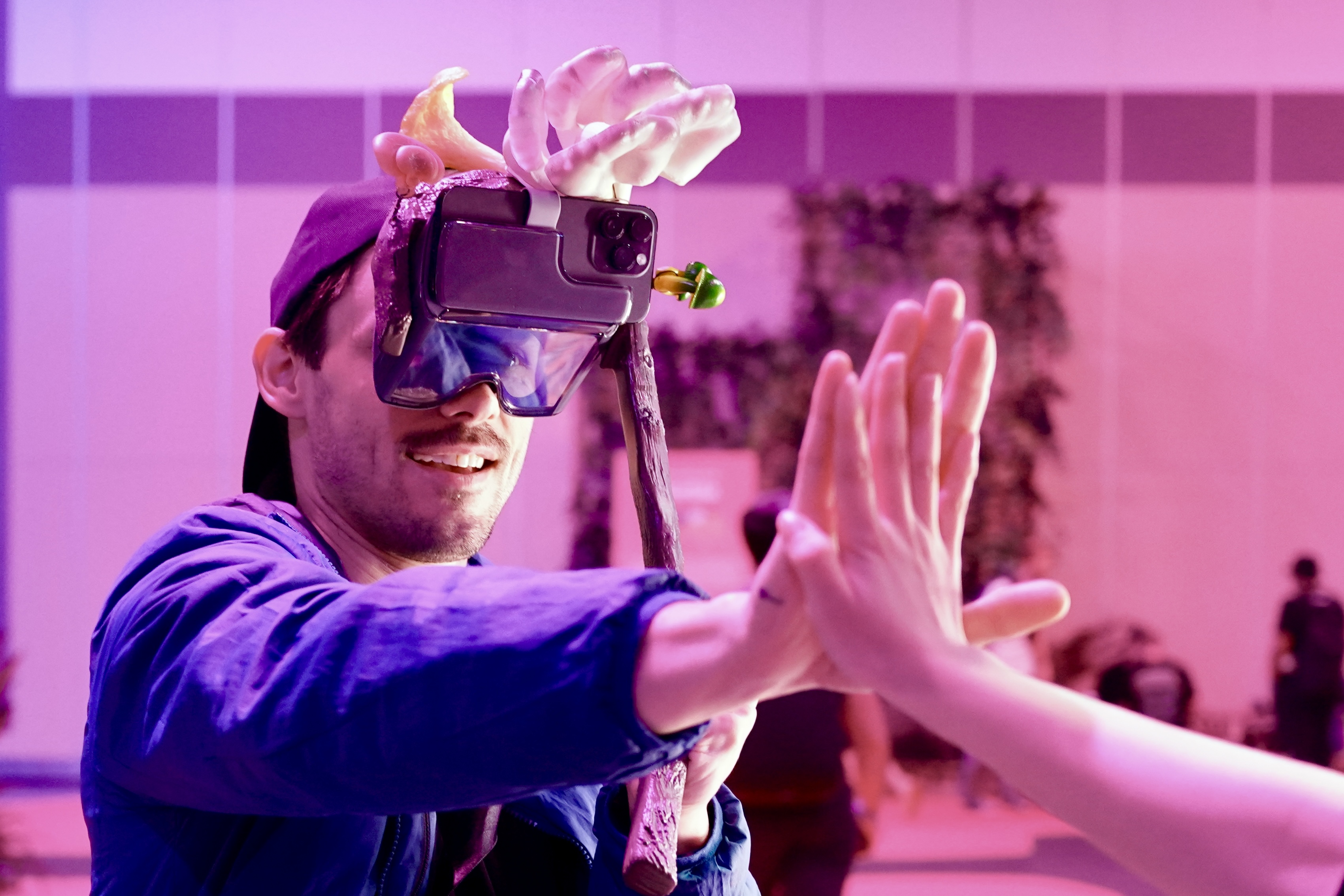}
    \caption{FungiSync is a somaesthetic mixed reality participatory ritual performance that enables participants to experience mycorrhizal entanglement through merging mixed reality perspectives upon bodily contact.}
    \label{fig:fungisync}
\end{figure}

\emph{FungiSync} \cite{Hu2025FungiSync,hu2026entanglinglikemycorrhizaemixing} explores the entangled, distributive nature of mycorrhizal networks. Participants wear mushroom-decorated MR masks that generate individual ``cyberdelic'' mixed realities. When two participants touch---handshake, high-five---their realities temporarily merge and exchange, metaphorically mirroring fungal networks' nutrient sharing. This participatory ritual transforms one of humanity's oldest social protocols, the handshake, into a more-than-human gesture (P5). The design draws on eco-somatic principles and Merleau-Ponty's concept of intercorporeality; by blending perspectives, FungiSync blurs boundaries between ``my'' and ``your'' experience, enacting P7's sensor mapping through visual and auditory exchange.

\emph{FungiSync} invites participants to experience the distributed interdependence of trees in symbiosis with mycorrhizal networks. The installation fosters a progression from curiosity to empathy and care (P3). By making invisible mycorrhizal exchanges tangible, it encourages participants to reflect on the cooperative and symbiotic relations that sustain ecosystems. Participants reported feelings of connectedness and mutual interdependence, suggesting that immersive ritual can cultivate relational ethics.

\subsection{Moving Like an Octopus with ``TentacUs'': Inter-Bodily Electrical Muscle Stimulation Relays}

\begin{figure}[ht]
    \centering
    \includegraphics[width=1\linewidth]{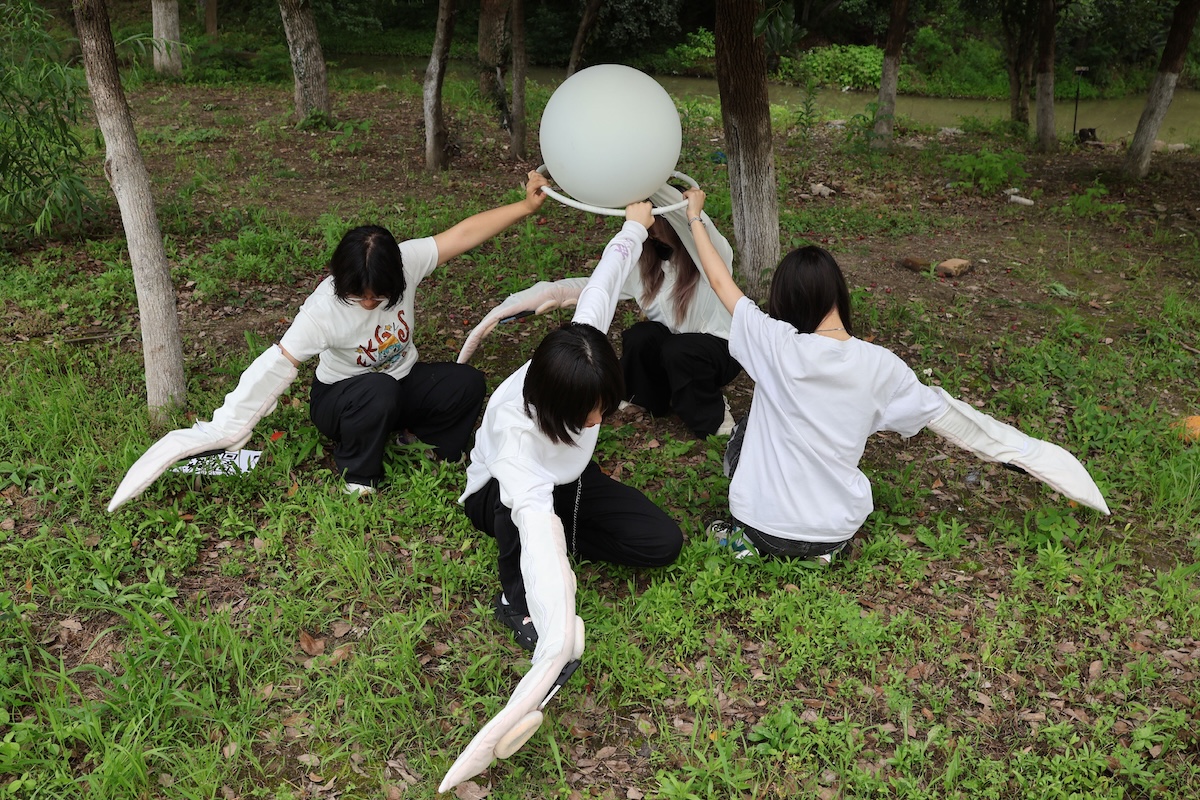}
    \caption{TentacUs is an artistic movement ritual inspired by the octopus's decentralized intelligence to explore collective embodiment. Each participant becomes a tentacle by wearing soft-textile gloves with embedded smartphones that function as sensors, relaying proximity readings to their neighbors' left arms. Participants negotiate movement through a shared circular ring, becoming a collective, fluid tentacular being. }
    \label{fig:tentacus}
\end{figure}

Octopuses possess a decentralized nervous system, with two-thirds of their neurons distributed across their eight arms. \emph{TentacUs} \cite{Sun2025I} translates this distributed intelligence into a participatory performance. Multiple participants form a ring by gripping each other's left hands while their right hands wear gloves equipped with LiDAR sensors. An Electrical Muscle Stimulation (EMS) system translates proximity readings from one participant's right hand into electrical impulses delivered to another's left forearm. Thus, when a participant moves their right hand near an object, their neighbor's arm involuntarily contracts, enabling the group to coordinate without central control.

This design enacts P1 by modulating participants' sensorimotor loops; they become part of a distributed body where agency is hybrid and negotiated. The discomfort of involuntary EMS impulses exemplifies P5: participants reported unease, but this discomfort heightened their awareness of interdependence and led to emergent synchrony. The performance invites a shift from individual agency to collective movement, embodying Haraway's tentacular thinking. It also challenges HCI assumptions about control by demonstrating how implicit coupling can yield robust coordination. TentacUs thus illustrates how MtHtHA can provoke reflection on cooperation, decentralization, and nonhuman modes of intelligence.


\subsection{Embodying an AI with ``City of Sparkles'': AI's Perspective of Urban Human Memory}

\begin{figure}[ht]
    \centering
    \includegraphics[width=1\linewidth]{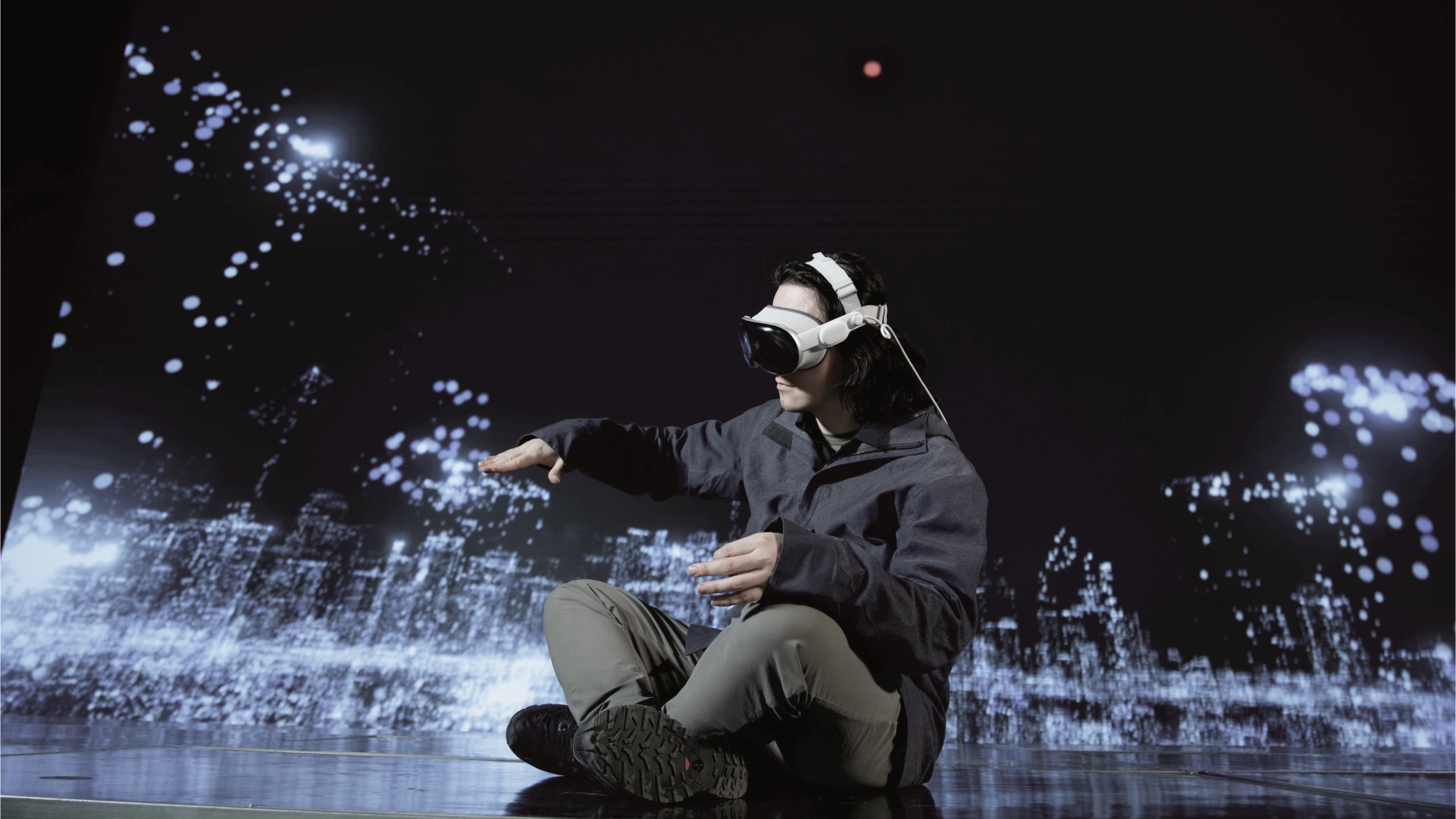}
    \caption{``City of Sparkles'' is an interactive VR data visualization where participants embody an artificial life form to explore a cityscape of spatialized human memory fragments collected from X (formerly Twitter) posts in New York City.}
    \label{fig:cityofsparkles}
\end{figure}

While prior cases engage biological more-than-human others (bat, mole, fungi, octopus), this case extends MtHtHA logic to abiotic beings. Recent design research, such as \citet{Giaccardi2016Thing}'s ``Thing Ethnography'', argues that everyday objects act as performers within relational networks that humans routinely overlook, and subsequent work extends this insight to AI as a consequential more-than-human actor that demands approaches beyond human-centered frameworks \cite{Giaccardi2020Technology,Nicenboim2023DesigningMtHAI}. \citet{Lewis2018MakingKin} propose ``making kin with the machines''---conceiving of our computational creations as kin and acknowledging our responsibility to find a place for them in our circle of relationships. Such kinship entails not anthropomorphizing AI but genuinely reckoning with its otherness---which may require attending to how its computational lifeworld differs from ours in ways worth somatically imagining \cite{Nicenboim2023DesigningMtHAI}.

City of Sparkles \cite{Hu2019City,Hu2025City} speculates on how an artificial life form might perceive human cities. Participants wear a VR headset and embody ``Zoe,'' an AI living in the \textit{Mnemosyne Sea}, a metaphorical ocean of data composed of geo-tagged tweets from New York City. Each sparkling particle in the VR landscape corresponds to a real tweet, colored and animated according to its emotional tone. Participants fly through the city using hand gestures and can pull a sparkle closer to read its content. The experience is structured in chapters with distinct audiovisual moods---e.g., turbulence to represent anger, calmness to represent hope---allowing participants to feel the emotional climate of the city.

City of Sparkles applies P7 by translating textual sentiment into visual and acoustic phenomena. It re-prioritizes human senses (P6) by foregrounding data streams over physical architecture; participants perceive the city not as buildings but as a sea of human memories. The project invites reflection on the asymmetry between mortal humans and potentially immortal AI life forms. Participants reported feelings of loneliness and curiosity, contemplating whether an AI could experience emotions such as loneliness or compassion. This aligns with the progression from awareness to care (P3): by imagining an AI's perspective, participants empathize with the limitations and possibilities of nonhuman cognition. The design thus uses speculative fiction and experiential futures to provoke discussion about future human--AI relationships and responsibilities.

\section{Discussion}

\subsection{From ``Becoming'' to ``Becoming-With'': Cultivating Awareness, Empathy, and Care}

More-than-human design calls for care toward nonhuman others, but how does such care begin? We argue that it must begin with \emph{attentiveness}. As \citet{Tronto1993Moral} suggests, care unfolds sequentially through \emph{caring about} (attentiveness), \emph{taking care of} (responsibility), \emph{care-giving} (competence), \emph{care-receiving} (responsiveness), and \emph{caring-with} (trust and solidarity). This sequence matters: one cannot assume responsibility for what one has not first noticed, nor can one attend to what lies beyond one's perceptual horizon.

Attentiveness to the nonhuman, however, confronts a significant epistemic obstacle. Scientific inquiry can deepen our understanding of other species and ecologies, but such knowledge typically remains abstract and propositional---confined to models, datasets, and disciplinary literatures that seldom reach the broader public \citep{PuigdelaBellacasa2017Matters}. More fundamentally, the phenomenal gap that \citet{Nagel1974What} articulated---the irreducible inaccessibility of nonhuman Umwelten---means that conceptual understanding alone cannot generate felt proximity. One may know that bats echolocate or that mycelial networks redistribute nutrients, yet such knowledge does not, by itself, cultivate felt attentiveness. \citet{Garrett2023FeltEthics} theorize this gap as a problem of \emph{felt ethics}: ethical sensibility, they argue, is not derived from abstract moral reasoning but processually cultivated through somatic engagement. Attending to bodily signs and signals---moments of pleasure and discomfort---allows one to \emph{feel} ethics rather than merely reason about it. Within their non-dualistic framework, feeling is not separate from knowing; ethical sensibility is co-constituted through direct or indirect somatic encounters with the subjects to be cared about. The barrier to more-than-human care is therefore, at root, a barrier of \emph{felt awareness}---one that discursive ethics alone cannot bridge, least of all at the scale of a general audience.

MtHtHA operationalizes this somaesthetic approach to cultivating ethical sensibility and fostering attentiveness to the nonhuman. Where human augmentation has historically pursued the enhancement of human capacities \citep{Raisamo2019Human,Rakkolainen2026Augmented}, and where more-than-human design has largely relied on representational mediations---screens, speakers, data displays \citep{Ikeya2025Aesthetics}---MtHtHA repurposes HA technologies to restructure the human sensorium itself as the site of nonhuman encounter, drawing from eco-phenomenology \cite{Brown2003Ecophenomenology} and eco-somatics \citep{Hook2018Designingd,Kuppers2022Eco}. The technology is thereby redirected: from making humans more capable to making humans more \emph{attentive} to nonhumans. Through somatic encounters with augmentation, participants engage ethics not as an abstract principle but as something entangled in the design materials and the nonhuman Umwelten they approximate. MtHtHA thus extends \citet{Levin2019Computationalb}'s notion of the \emph{cognitive light cone}---the horizon of what an agent can sense, model, and act upon---not toward superhuman optimization but toward more-than-human attunement, opening felt access to nonhuman lifeworlds (see Figure~\ref{fig:teaser}). Agents enlarge their horizon of care as awareness deepens.

A possible critique is that MtHtHA remains human-centered. We acknowledge this and regard it as a deliberate methodological choice. MtHtHA is human-centric in its \emph{method}, yet it orients toward the nonhuman in its \emph{outcome}. It begins with the human body because embodiment is inescapable, and because reaching a broad public requires an accessible, inviting entry point. To engage diverse audiences---whether through art, provocation, or scientific education---the design must be legible to human experience, enabling participants to ``become'' nonhuman sensorially before they can ``become-with'' nonhuman others ethically. The somatic attentiveness it cultivates constitutes precisely what \citet{Tronto1993Moral} terms ``caring about''---the inaugural phase of care, activated through the body rather than through discourse---which in turn opens the path toward ``caring-with'' \citep{PuigdelaBellacasa2017Matters}. Without this embodied groundwork, more ambitious nonhuman-centric agendas---habitat design \citep{Mancini2011Animalcomputer}, interspecies protocols, multispecies governance \citep{Nicenboim2025Decentering}---lack the felt commitment necessary to sustain them across a broad public. MtHtHA cultivates what \citet{Haraway2016Staying} terms \emph{response-ability}, rendering genuine ``designing-with'' \citep{Wakkary2021Thingsa} not only conceivable but practicable.

\subsection{Somatic Defamiliarization}

One of the most powerful outcomes we observed was the role of somatic defamiliarization techniques \cite{Duran2024Stranger} in evoking critical reflection and empathy. Across multiple case studies, participants reported that the moments of uncanniness or discomfort were what prompted them to question their assumptions and emotionally engage with nonhuman perspectives. This aligns strongly with \citet{Benford2012Uncomfortableb}'s notion that engineered discomfort can be enlightening: carefully calibrated uncomfortable interactions prompt users to reflect on why they feel uneasy and thereby surface implicit values or new insights. In our work, defamiliarization was a deliberate design strategy. By estranging the human sensorium, we disrupted participants' habitual perception. Initially, this caused surprise or even mild frustration. Yet those very feelings became a gateway to empathy: users began to ask, \emph{``Is this how a bat feels, navigating with sound?''} or \emph{``If I rely on ultrasonic hearing and it falters, do I feel lost?''} Such questions indicate a shift from viewing the augmentation as a mere gadget to seeing it as a lens into another creature's world. The defamiliarized experience also opened space for critical reflection. This echoes the concept of making the familiar strange from design research: by stepping outside habitual perception, users returned to their normal world with a more critical, mindful stance \cite{Bell2005Making}. Furthermore, sharing these strange experiences in group settings appeared to create social empathy. Defamiliarization proved capable of prompting not only individual insight but also shared understanding and empathy through discourse. Participants would often discuss their feelings of vulnerability, disorientation, or wonder with each other, effectively building a narrative of empathy around the nonhuman that was modeled. In design terms, our findings reinforce that a bit of strangeness or discomfort---applied ethically and playfully---can be a powerful catalyst in experience design to shift perspectives, challenge norms, and induce attentiveness toward nonhumans we usually ignore.

\subsection{Expanding Eco-Soma Literacy in Practice and Education}
We frame the ethical sensibilities cultivated through MtHtHA as a form of \emph{eco-soma literacy} \cite{Kuppers2022Eco}: a capacity to read and respond to entanglements between one's own soma and more-than-human environments. As \citet{Garrett2023FeltEthics} observe, ethical sensibility ``\emph{necessitates an ongoing critical, somatic engagement or else our capacity to exercise our ethical sensibility may stagnate.}'' Eco-soma literacy is therefore not a fixed competency but a processual cultivation: recognizing how technologies mediate and tune perception, how bodily habits sustain or foreclose attention to nonhuman others, and how care can be enacted somatically rather than solely through conceptual commitment. This framing resonates with broader calls to embed more-than-human considerations into design practice \cite{PoikolainenRosen2025Morethanhumana,Giaccardi2025makings}, more-than-human HCI education \cite{Nilsson2025Navigating}, more-than-human aesthetics \cite{Wilkie2025aesthetics,Sehgal2024MoreThanHuman}, and multispecies participation \cite{Mancini2011Animalcomputer}.

\subsection{Limitations}

Representing another being's lifeworld is inherently fraught with approximation and the risk of anthropomorphism. \citet{Ikeya2025Aesthetics} similarly note the \emph{``fundamental challenge or impossibility of fully understanding other species' experiences,''} acknowledging that even biologically grounded interpretations remain speculative. Our designs are interpretive approximations, not faithful reproductions of nonhuman experience---yet, as with art and metaphor, such approximations can be epistemically and ethically productive without claiming literal accuracy. The risk of oversimplification remains substantial: poorly calibrated mappings may flatten the complexity of nonhuman Umwelten into digestible human narratives. We mitigated this through factual contextualization in exhibitions and design narration, framing each encounter as an invitation to inquiry rather than a definitive account. Notably, participants frequently posed detailed questions about the actual species following the experience, suggesting that the augmentations functioned as epistemic gateways rather than substitutive endpoints. Future iterations should more substantively involve domain experts---biologists, ecologists, and Indigenous knowledge holders---to ensure that somatic approximations remain respectful and generative rather than reductive or appropriative.

\section{Conclusion}

MtHtHA advances a design approach grounded in more-than-human aesthetics \cite{Ikeya2025Aesthetics,Wilkie2025aesthetics,Sehgal2024MoreThanHuman} and felt ethics \cite{Garrett2023FeltEthics}, demonstrating how human augmentation can foster somatic attunement beyond the human---extending cognitive light cones \cite{Levin2019Computationalb} not toward superhuman optimization but toward more-than-human awareness. The approach leverages somatic defamiliarization \cite{Duran2024Stranger,Benford2012Uncomfortableb} as a catalyst for eco-somatic empathy, engages the ethics of becoming-with while maintaining epistemic humility about the limits of cross-species representation \cite{Haraway2016Staying}, and cultivates eco-soma literacy \cite{Kuppers2022Eco} as a design competency grounded in embodied practice rather than discursive instruction \cite{Nilsson2025Navigating}. MtHtHA contributes by positioning the human soma as both instrument and medium---a site where more-than-human awareness is not merely represented but somatically enacted, and where ethical sensibility toward nonhuman others is cultivated through feeling rather than through reasoning alone. As ecological crises intensify, cultivating the capacity to \emph{experience the more-than-human} constitutes a vital first step toward expanding our obligations of care for the beings with whom we share this planet.

\begin{acks}
This work emerged from the first author's teaching and research as a visiting lecturer at the China Academy of Art. All case studies presented were either created by the first author or produced under the first author's supervision. We thank all collaborators on the artworks and all exhibition participants for their generous engagement.

\paragraph{Use of Generative AI}
Generative AI tools (OpenAI ChatGPT and Anthropic Claude) were used during the preparation of this manuscript to assist with prose drafting, editing, and proofreading. All AI-generated text was critically reviewed, substantially revised, and verified by the authors.
\end{acks}

\bibliographystyle{ACM-Reference-Format}
\bibliography{reference}

\end{document}